\documentclass[11pt,twoside,a4paper]{article}
\usepackage[affil-it]{authblk}
\usepackage{ amssymb }
\usepackage{amsmath}
\usepackage{tensor}
\usepackage{color}
\usepackage{tikz}
\usetikzlibrary{matrix}
\usepackage[margin=2.9 cm]{geometry}
\usepackage{polski}
\usepackage[cp1250]{inputenc}
\usepackage[polish,english]{babel}
\usepackage[toc,page]{appendix}

\input xy
\xyoption{all} \tolerance=500

\def\sT{\mathsf T}
\newdir{ (}{{}*!/-5pt/@^{(}}
\newdir{|>}{!/4,5pt/@{|}*:(1,-.2)@^{>}*:(1,+.2)@_{>}}

\def\mg{\mathfrak g}
\def\d{\mathrm d}
\def\TM{\mathsf TM}
\def\sV{\mathsf V}

\def\sJ{\mathsf J^1}
\def\sJd{\mathsf J^\dagger}
\def\sJo{\mathsf J^o}
\def\J2{\mathsf J^2}
\def\sj{\mathsf j^1}
\def\ind{\indices}
\def\blue{\color{blue}}
\def\red{\color{red}}

\def\Ad{\mathrm{Ad}}
\def\ad{\mathrm{ad}}

\def\ddt{\frac{\mathsf d}{\mathsf dt}_{|t=0}}

\def\wdt{\widetilde }
\def\bh{\textbf{h}}

\newtheorem{theorem}{Theorem}

\title{Hamilton-Jacobi theory for gauge field theories}
\author{Manuel de Le\'on}
\affil{Instituto de Ciencias Matemáticas (CSIC-UAMUC3M-
UCM), c\textbackslash Nicolás Cabrera, 13-15, Campus Cantoblanco, UAM
28049 Madrid, Spain
and
Real Academia de Ciencias Exactas, Físicas y Naturales, c\textbackslash  de Valverde,
22, 28004 Madrid, Spain
E-mail address: mdeleon@icmat.es}
\author{Marcin Zając}
\affil{Department of Mathematical Methods in Physics, University of Warsaw, Pasteura 5, 02-093, 
Warsaw, Poland}

\begin{document}
\maketitle

\begin{center}
\textbf{Abstract.}
Recently, M. de Le\'on {\it el al.} (\cite{HJ1}) have developed a geometrical description of Hamilton-Jacobi theory for multisymplectic field theory. In our paper we analyse in the same spirit a special kind of field theories which are gauge field theories. The Hamilton-Jacobi theory for this kind of fields is shown.

\end{center}

\tableofcontents

\section{Introduction}\label{sec:1} 

The historical beginning of classical field theory comes to XIX century with the discovery of electromagnetic field by James Clerk Maxwell. However only in XX century physicists and mathematicians started to look for a proper mathematical description of phenomena described by this theory (see e.g. \cite{F1,F2,F3,F4,F5,F6,F7,F8,F9,F10,F11}). From the mathematical point of view a classical field is represented by a section of the fiber bundle $\pi:E\to M$ where the base manifold $M$ usually represents spacetime. For instance a real (complex) scalar field is a section of the trivial bundle $M\times\mathbb R\to M$ ($M\times\mathbb C\to M$), electromagnetic field is a section of the cotangent bundle $\sT^*M\to M$ etc. The natural choice for the configuration space of the first order classical field theory is a bundle of first jets $\sJ E$ of sections of $\pi$. The bundle $\sJ E$ plays a similar role in field theory as the tangent bundle $\sT M$ in classical mechanics, however, one has to be carefull with this comparisement since the internal structure of $\sJ E$ is much richer than $\sT M$.

In mechanics one uses symplectic geometry to derive dynamics of the mechanical system. The cotangent bundle $\sT^*M$ associated with the Hamiltonian description of the system has a canonical structure of the symplectic manifold. On the other hand, the bundle $\sT M$ does not have a canonical symplectic form but, at least for regular systems, one can transport symplectic structure from $\sT^*M$ to $\sT M$ via Legendre map \cite{Klein}. In field theory one has to consider a multisymplectic geometry which is a generalisation of the symplectic structure. It turns out that in the presence of a volume form on the $(n+1)$-dimensional manifold $M$, the phase bundle is isomorphic to the bundle $\Lambda^{n+1}_2E$ representing two-horizontal $(n+1)$-forms on $E$ and being a canonical mutlisymplectic manifold. To read more about multisymplctic structure and its applications in field theory one can see for example \cite{MP1,MP2,MP3,MP4,M5,M6}.

In the very center of interest of physicists are so called {\it gauge theories}. Gauge theories are field theories with certain special kind of symmetries called {\it gauge transformations}. In the modern description of fundamental interactions one usually takes a classical Yang-Mills theory and then quantize it to obtain possible experimental predictions.

On the other hand Hamilton-Jacobi theory is a powerful tool in analitycal mechanics that provides a method to find the dynamics of the mechanical system. It also the most clear way to see the classical limit of the quantum mechanics when one takes $\hslash\to 0$. Its geometrical formulation was investigated in many papes, e.g. in {\cite{HJ3,HJ5,HJ6,HJ7,HJ8,HJ9,HJ10,HJ11,HJ12,HJ13,HJ14,HJ15,HJ17,HJ18} }. Recently a Hamilton-Jacobi for field theories version has been developed too \cite{HJ1,HJ2,HJ4,HJ16}. In our paper we would like to investigate the field theoretical version of Hamilton-Jacobi theory for gauge field theories.   

The paper is organised as follows. In section \ref{Multisymp} we present a basic features of the geometrical picture of classical field theory of first order. We start with a general definition of a multisymplectic structure. Subsequently, we present basic tools of jet bundle theory and their relations to multisymplectic structures. In \ref{princip} we make a brief introduction to principal bundles which provide the most common definition of a gauge field. The main result from this constructions is that a gauge field is just a connection in a principal bundle. In section \ref{gaugejet} we present an equivalent definition of connection which is based on jet bundles and is more useful for our purposes. In particular we present a canonical projection of the configuration space of gauge field theory which is essential in our further work. Section 5 contains Hamiltonian formalism for gauge theories. In section we present the main result of our paper which is a Hamilton-Jacobi theory for gauge theories.

In our paper we will assume that the base manifold $M$ has always dimension $n+1$. If $(x^i)$ are local coordinates in $M$ we introduce notation
$$\d^{n+1}x=\d^{0}x\wedge...\wedge\d^{n}x   $$
and
$$\d^{n}x_i=i_{\frac{\partial}{\partial x^i}}\d^{0}x\wedge...\wedge\d^{n}x   $$
for the contraction with the coordinate vector fields. Furthermore, we assume that $M$ is orientable with fixed orientation, together with a determined volume form $\eta$. Its pullback to any bundle over $M$ will still be denoted $\eta$, as for instance $\pi^*\eta$. In addition, local coordinates on $M$ will be chosen compatible with $\eta$, which means such that $\eta=\d^{n+1}x$. To great extent, this form $\eta$ is not needed and our constructions can be generalized, although we are going to make use $\eta$ for the sake of simplicity.

\section{Multisymplectic structures and Jet bundles}\label{Multisymp}
We begin reviewing the basic notions of multisymplectic structures and jet bundles. 

\subsection{Multisymplectic structures}

Let $V$ be finite-dimensional real vector space. A $(k+1)$-form $\Omega$ on $V$ is said to be multisymplectic if it is non-degenerate, i.e., if the linear map 
$$b_\Omega:V\to\Omega^kV^*  $$
$$v\longmapsto b_\Omega(v):=i_v\Omega  $$
is injective. The pair $(V,\Omega)$ is called then a {\it multisymplectic} vector space of order $k+1$. This definition has a straightforward extension to differential manifolds. We say that a pair $(P,\Omega)$ where $P$ is a manifold and $\Omega$ a closed $(k+1)$-form on $P$ is a multisymplectic manifold if each $(\sT_pP,\Omega(p))$ is multisymplectic.

Same as the contangent bundle is a canonical example of symplectic manifold, the canonical example of a multisymplectic manifold is a bundle of forms $P=\Lambda^kN$ over a manifold $N$.

Let $N$ be a smooth manifold of a dimension $n$, $\Lambda^kN$ be the bundle of $k$-forms on $N$ and $\nu:\Lambda^kN\to N$ be the canonical projection $(1\le k\le n)$. The {\it Louville} or {\it tautological form} onf order $k$ is the $k$-form $\Theta$ over $\Lambda^kN$ given by

$$\Theta(\omega)(v_1,...,v_k)=\omega(\sT_\omega\nu(v_1),...,\sT_\omega\nu(v_k)), \quad \omega\in\Lambda^kN, \quad v_1,...,v_k\in\sT_\omega(\Lambda^kN).$$
The canonical multisymplectic form is then defined by

$$\Omega=-\d\Theta.  $$ 
Let us introduce local coordinates $(x^i)$ in $N$ and induced local coordinates $(x^i, p_{i_1...i_k})$ in $\Lambda^kN$ where $1\leq i_1\leq ...\leq i_k\leq n$. Then 
$$\Theta=\sum\limits_{i_1<i_2<...<i_k}p_{i_1...i_k}\d x^{i_1}\wedge...\d x^{i_k} $$
$$\Omega=-\sum\limits_{i_1<i_2<...<i_k}\d p_{i_1...i_k}\wedge\d x^{i_1}\wedge...\d x^{i_k}.  $$
It is immediate to check that $\Omega$ is indeed multisymplectic. 

Another example of multisymplectic manifold comes from the particular case of the previous example. Let $\pi:E\to M$ be a bundle. Let us assume that $\dim M=n+1$ and $\dim E=n+1+m$. Given $1\le r\le m$, we can consider the vector subbundle $\Lambda^k_r E$ of $\Lambda^k E$ that consists of $k$-forms on $E$ which are $r$-horizontal with respect to the fibration $\pi$, i.e.
$$(\Lambda^k_r E)_e=\{\omega\in\Lambda^k_eE:i_{v_r}...i_{v_1}\omega=0, \quad \forall v_1,...,v_r\in\sV_eE   \}.  $$
where $e\in E$ and $\sV E\subset\sT E$ is a subbundle of vectors  tangent to $E$ which are vertical with respect to $\pi$. 

We denote by $\nu_r,\Theta_r, \Omega_r$ restrictions to $\Lambda^k_r E$ of $\nu,\Theta, \Omega$ respectively. It is easy to see that $(\Lambda^k_r E,\Omega_r)$ is a multisymplectic manifold. The most interesting cases from the point of view of field theory are $k=n+1$ and $r=1,2$.

Let $(x^i,u^\alpha)$ denote adapted coordinates on $E$, where $i\in\overline{0,n}$ and $\alpha\in\overline{1,m}$. They induce coordinates $(x^i,u^\alpha,p,p\ind{^i_\alpha})$ on $\Lambda^{n+1}_2E$ such that any element $\omega\in\Lambda^{n+1}_2E$ has a form 
$$\omega=p\d^{n+1}x+p\ind{^i_\alpha}\d u^\alpha\wedge\d^nx_i,   $$
where $\d^nx_i=i_{\frac{\partial}{\partial x^i}}\d^{n+1}x$. Therefore $\Theta_2$ and $\Omega_2$ in local coordinates read

$$\Theta_2=p\d^{n+1}x+p\ind{^i_\alpha}\d u^\alpha\wedge\d^nx_i$$
$$\Omega=-\d p\wedge\d^{n+1}x-\d p\ind{^i_\alpha}\wedge\d u{^\alpha}\wedge\d^{n}x_i.$$

\subsection{First order jet bundles} 

We will introduce now the notion of first order jet spaces and their duals. We will base here on a notation from \cite{KG}. For more detailed discussion of the jet bundle geometry see e.g. \cite{Saunders}.

Let $\pi :E\to M$ be a bundle with the total space of the dimension $\mathrm{dim}  E=n+1+m$. We introduce in a domain $U\in M$ a local coordinate system $(x^i)^{n}_{i=0}$ on $M$. In field theory, fields are represented by sections of the fibration $\pi$. The total space is a space of values of the field e.g vector fields are sections of the $\pi$ being a vector bundle, scalar fields are sections of the trivial bundle $E=M\times\mathbb R$ or $E=M\times\mathbb C$, etc. On an open subset $V\subset E$ such that $\pi(V)=U$ we can introduce local coordinates $(x^i,u\ind{^\alpha})$ adapted to the structure of the bundle.

In $\sT E$ we have a vector subbundle $\sV E$ of the vectors vertical with respect to the projection $\zeta$ i.e. $\sT\pi(v_p)=0$ for $v_p\in\sV_pE$. We will also need the dual vector bundle $\sV^*E$.

The space of first jets of sections of the bundle $\pi$ will be denoted by $\sJ E$.  By definition, the first jet
$\sj_m\phi$ of the section $\phi$ at the point $m\in M$ is an equivalence class of sections having the same value at the point $m$ and such that the spaces tangent to the graphs of the sections at the point $\phi(m)$ coincide. Therefore, there is a natural projection $\sj\pi$ from the space $\sJ E$ onto the manifold $E$

$$\mathsf j^1\pi:\sJ E\to E: \quad  \mathsf j^1_m\phi\longmapsto\phi(m).    $$
Moreover, every jet $\sj_m\phi$ may be identified with a linear map $\sT\phi:\sT_mM\to\sT_{\phi(m)}E$. Linear maps coming from jets at the point $m$ form an affine subspace in a vector space $Lin(\sT_mM,\sT_eE)$ of all linear maps from $\sT_mM$ to $\sT_eE$. A map belongs to this subspace if composed with $\sT\pi$ gives identity. In a tensorial representation we have an inclusion
$$\sJ_eE\subset\sT^*_mM\otimes\sT_eE.$$ 
It is easy to check that the affine space $\sJ_eE$ is modelled on a bundle $\sT_m^*M\otimes\sV_eE$. Summarising, the bundle $\sJ E\to E$ is an affine subbundle in a vector bundle
$$\pi^*(\sT^*M)\otimes_E\sT E\to E$$   
modelled on a vector bundle 
$$\pi^*(\sT^*M)\otimes_E\sV E\to E.$$
The symbol $\pi^*(\sT^*M)$ denotes the pullback of the cotangent bundle $\sT^*M$ along to the projection $\zeta$. In the following we will omit the symbol of the pullback writing simply $\sT^*M\otimes_E\sT E$ and $\sT^*M\otimes_E\sT E$.

Using the adapted coordinates $(x^i,u\ind{^\alpha})$ in $V\subset E$, we can construct the induced coordinate system $(x^i,u\ind{^\alpha},u\ind{^\beta_j})$ on $\mathsf j^1\pi^{-1}(V)$ such that for any section $\phi$ given by $n$ functions $\phi^a(x^i)$ we have

$$u\ind{^\beta_j}(\phi^\alpha(x^i))=\frac{\partial\phi^\beta }{\partial x^j}(x^i(m)).$$
In the tensorial representation the jet $\mathsf j^1_m\phi$ may be written as
$$\d x^i\otimes\frac{\partial}{\partial x^i}+\frac{\partial\phi^\alpha}{\partial x^j }(x^i(m))\d x^j\otimes\frac{\partial}{\partial u\ind{^\alpha}},    $$
where we have used local bases of section of $\sT^*M$ and $\sT E$ coming from the chosen coordinates.

We will introduce now the bundle which is dual to the bundle $\sJ E\to E$. Let us recall that each fiber $\sJ_eE$ is an affine space. We can consider a set of affine maps $\sJ_eE\to\mathbb R$ which we will denote by $\mathsf{Aff}(\sJ_eE,\mathbb R)$, for each $e\in E$. Collecting $\mathsf{Aff}(\sJ_eE,\mathbb R)$ point by point we obtain a bundle of affine maps on $\sJ E$, namely $\mathsf{Aff}(\sJ E,\mathbb R)\to E$. From now we will use notation $\sJd E:=\mathsf{Aff}(\sJ E,\mathbb R)$. It is a vector bundle over $E$. If $(x^i,u\ind{^\alpha},u\ind{^\beta_j})$ are coordinates in $\sJ E$ then we introduce coordinates $(x^i,u\ind{^\alpha},r,\varphi\ind{^b_j})$ in $\sJd E$. The evaluation between $\sJ E$ and $\sJd E$ in coordinates reads
$$\sJd E\times_E\sJ E\to\mathbb R, \qquad \langle T_e,\mathsf j_e\psi\rangle=r+\varphi\ind{^b_j}y\ind{^j_b}   $$
We also introduce a vector bundle $\mathsf J^oE$ over $E$ which is a quotient of $\sJd E$ by constant affine maps, namely
$$\mathsf J^oE:=\sJd E/\{f:E\to\mathbb R\}.$$
It is equipped with adapted coordinates $(x^i,u\ind{^\alpha},\varphi\ind{^b_j})$. The bundle $\mu:\sJd E\to\sJo E$ is a principal $\mathbb R$-bundle.

The following theorem justifies the importance of the multisymplecitc structures in field theory.
\begin{theorem}
There exists an isomorphism $\Psi:\Lambda^{n+1}_2 E\to\sJd E$ given by the formula
$$\langle\Psi(\omega),\sj_x\phi\rangle\eta=\phi_x^*(\omega),\qquad  \forall\sj_x\phi\in\sJ_xE,\quad \forall \omega\in\Lambda^{n+1}_2 E.  $$
\end{theorem}
We will therefore identify $\sJd E$ with $\Lambda^{n+1}_2 E$ and $\sJo E$ with $\Lambda^{n+1}_2 E/\Lambda^{n+1}_1 E$.

\section{Geometry of principal bundles}\label{princip} 

We introduce now a mathematical framework to analyse gauge theories. It turns out that gauge fields are sections of the bundle of connections in principal bundle. Therefore we give main characteristics of the theory of principal bundles and connections in them.

\subsection{Principal bundles}\label{ssec:1}

Let $G$ be a Lie group and $P$ a smooth manifold. We will denote by $\mg$ a Lie algebra of $G$. We assume that $G$ acts on $P$ from the right-hand side in a smooth, free and proper way. We denote by $M$ a space of orbits $P/G$. The bundle $\pi:P\to M$ is called a {\it principal bundle}. It is locally isomorphic to $M\times G$. Let $\mathcal U_\alpha$ be an open subset in $M$. We have local trivialisations

$$\xymatrix{
\pi^{-1}U_\alpha  \ar[dr]_{\pi}  \ar[rr]^{\Psi_\alpha}   &  & U_\alpha\times G  \ar[dl]^{pr_1}\\
   & U_\alpha & 
}
$$
where $\Psi_\alpha$ is a $G$-equivariant diffeomorphism such that $\Psi_\alpha(p)=(\pi(p),g_\alpha(p))$ for $g_\alpha:P\to G$. Equivariance means that $\Psi_\alpha(pg)=\Psi_\alpha(p)g$ which implies that $g_\alpha$ is also $G$-equivariant in a sense  $g_\alpha(pg)=g_\alpha(p)g$. 

Local trivialisations of principal bundles are associated with its local sections. Let $\sigma_\alpha: U_\alpha\to\pi^{-1}U_\alpha$ be a local section of $\pi$. Using $\sigma_\alpha$ we can construct local trivialisation by $\Psi_\alpha(\sigma_\alpha(m))=(\pi(\sigma_\alpha(m)),e)$, where $e$ is the neutral element of $G$. Since $\Psi_\alpha$ is equivariant we obtain a trivialisation of the whole subset $\pi^{-1}U_\alpha$. Conversely given a trivialisation $\Psi_\alpha$ we define value of $\sigma_\alpha(m)$ by condition $\Psi_\alpha(\sigma_\alpha(m))=(\pi(\sigma_\alpha(m)),e)$. Let us notice that $g_\alpha\circ\sigma_\alpha(m)=e$ for each $m\in M$. Local trivialisations (or equivalently local sections) enable us to identify fibers of the bundle $\pi$ with a group $G$ by choosing an element $p$ which satisfies $g_\alpha(p)=e$. 

Let $U_{\alpha\beta}:=U_\alpha\cap U_\beta$. For $\Psi_\alpha,\Psi_\beta$ associated with $U_\alpha$ and $U_\beta$ respectively we have transition conditions such that the following diagram commutes
$$\xymatrix{
U_{\alpha\beta}\times G  \ar[dr]_{pr_1} & \pi^{-1}U_{\alpha\beta} \ar[l]_{\Psi_\alpha}  \ar[r]^{\Psi_\beta}   \ar[d]^{\pi}  & U_{\alpha\beta}\times G  \ar[dl]^{pr_1}\\
   & U_{\alpha\beta}  & 
}
$$
Let us assume now that $m$ belongs to $U_{\alpha\beta}$. With each trivialisation $\Psi_\alpha,\Psi_\beta$ there are associated functions $g_\alpha,g_\beta$. One can show that there exist a function $\bar g_{\alpha\beta}:P\to G$ such that
$$\bar g_{\alpha\beta}(p)=g_\alpha(p)g_\beta(p)^{-1}  $$ 
What is more, $\bar g_{\alpha\beta}$ is constant on each fiber i.e.  $\bar g_{\alpha\beta}(p)=\bar g_{\alpha\beta}(pg)$. Therefore, $\bar g_{\alpha\beta}$ defines a function 
$$ g_{\alpha\beta}:M\to G, \quad g_{\alpha\beta}(\pi(p)):=\bar g_{\alpha\beta}(p) $$
Transition functions $g_{\alpha\beta}$ satisfy cocycle conditions

\begin{eqnarray*}
g_{\alpha\beta}(m)g_{\beta\alpha}(m)=e &\text{on}&  U_\alpha\cap U_\beta \\
g_{\alpha\beta}(m)g_{\beta\gamma}(m)g_{\gamma\alpha}(m)=e     &\text{on}&   U_\alpha\cap U_\beta\cap U_\gamma .
\end{eqnarray*}

\subsection{Adjoint bundle}\label{sec:3}

Let $F$ be a manifold and let $G$ act on $F$ from the left hand side. We assume that $G$ acts on $P\times F$ by
$$ g(p,f)=(pg,g^{-1}f). $$
We denote by $N:=(P\times F)/G$ a space of orbits of this action. We have a bundle
$$\xi: N\to M,\quad  [(p,f)]\to \pi([p]), $$
which is called an {\it associated bundle} to a principal bundle $P$. Let $s_{\alpha}:M\supset\mathcal U_\alpha\to P$ be a local section of $P$ and let $N\supset\mathcal O_{\alpha}:=\xi^{-1}(\mathcal U_{\alpha})$. Then for each orbit $y\in\mathcal O_{\alpha}$, where $\xi(y)=m$, there exist a unique element $\chi_s(y)\in F$ such that $(s(m),\chi_s(y))$ belongs to $y$. A local section of $\pi$ provides then a local trivialization of $N$
$$N\supset\mathcal O_\alpha\to \mathcal U_\alpha\times F,\quad y\longmapsto (\xi(y),\chi_s(y)).$$

The most important examples of associated bundles to $P$ in context of our work are the bundles with fibers $F=\mg$ or $F=G$, i.e. $N=(P\times\mg)/G$ and $N=(P\times G)/G$. We will use the notation $\ad(P):=(P\times\mg)/G$ and $\Ad(P):=(P\times G)/G$. The action of $G$ on $\mg$ and $G$ on $G$ is an adjoint map given by
$$\Ad:G\times\mg\to\mg,\quad (g,X)\longmapsto\Ad_{g}(X),$$ 
$$\Ad:G\times G\to G,\quad (g,h)\longmapsto\Ad_{g}(h).   $$
respectively.

Let us take a closer look on a bundle $\xi:\ad(P)\to M$. It has a natural structure of the vector bundle with addition and multiplication given by
\begin{eqnarray}
[p,X_1]+[p,X_2]&=&[p,X_1+X_2]  \qquad p\in P,\quad X_1, X_2\in\mg  \\
\lambda[p,X]&=&[p,\lambda X] \qquad \lambda \in\mathbb R   
\end{eqnarray}

The "zero" vector in each fiber is an element $[p,0]$. The bundle $\ad P$ is called the {\it adjoint bundle}. Given a section $s_\alpha:U_\alpha\to P$ we have a local trivialisation of $\ad(P)$

$$ \ad(P) \to U_\alpha\times\mg, \qquad [p,X]\longmapsto \left(\pi(p),X_{s_\alpha}\right) \quad h\in G.$$ 
where $X_{s_\alpha}\in\mg$ is a unique element satisfying
$$[p,X]=[s_\alpha\left(\xi(p)\right),X_{s_\alpha}].$$ 
In above trivialisation the equivalence class is represented by its representative in $e\in G$ i.e. if in a local trivialisation $(p,X)=(\pi(p),g_\alpha(p),X)$ then $[p,X]\in\ad P$ in trivialisation reads $(\pi(p),\Ad_{g_\alpha(p)^{-1}}X)$.

The structure of $\ad P$ has its reflection in transition functions between different trivialisations of $P$. Every element $(p,X)$ is in relation with itself therefore
$$ (\pi(p),g_\alpha(p),X)\sim (\pi(p),g_\beta(p),X)  $$  
where $g_\alpha:\pi^{-1}U_\alpha\to G$ and $g_\beta:\pi^{-1}U_\beta\to G$ are two different trivialisations related by the transition function $g_{\alpha\beta}$. Since for each $p\in\pi^{-1}U_{\alpha\beta}$ the relation $g_\alpha(p)=g_{\alpha\beta}(\pi(p))g_\beta(p)$ is satisfied we obtain   
$$ (\pi(p),e,\Ad_{g_\alpha(p)^{-1}}\circ X)\sim (\pi(p),e,\Ad_{g_{\alpha\beta}(m)}\circ\Ad_{g_\alpha(p)^{-1}}\circ X).  $$
Therefore the local trivialisations of $\ad P$ must satisfy a condition
\begin{equation}\label{diag}
$$
\xymatrix{
 &\ad(P)  \ar[dl]_{\it{triv_1}}  \ar[dr]^{\it{triv_2}}&   &   &[p,X]  \ar[dl]_{\it{triv_1}}  \ar[dr]^{\it{triv_2}}&    \\
 U_{\alpha\beta}\times\mg \ar[rr] && U_{\alpha\beta}\times\mg                         &     (m,Y) \ar[rr]  && (m,\Ad_{g_{\alpha\beta}(m)}Y) 
}
$$
\end{equation}
on double overlaps $U_\alpha\cap U_\beta$.

\subsection{Forms with values in $\ad P$}

We are ready now to construct the bundle of forms on $M$ with values in $\ad P$. We will denote it by $\Omega^k(M,\mg)$ be a bundle of $\mg$-valued $k$-forms on $M$. We assume that $G$ acts on $\mg$ by
$$R_g^*X=\Ad_{g^{-1}}X, \qquad g\in G, \quad X\in\mg    $$

Let $\{\xi_\alpha\}$ be a set of local $k$-forms on $M$ such that for each $\alpha$, $\xi_\alpha\in\Omega^k(U_\alpha,\mg)$. We also require that for each double overlap $U_{\alpha\beta}$ the condition
\begin{equation}\label{ad}
\xi_\alpha(m)=\Ad_{g_{\alpha\beta}(m)}\circ\xi_\beta(m),   \qquad  m\in U_{\alpha\beta},\quad  g_{\alpha\beta}:U_{\alpha\beta}\to G  
\end{equation}
is satisfied. We claim that the family of forms $\{\xi_\alpha \}$ define a global section of the associated bundle $\Omega^k(M)\otimes\ad P$. Indeed, if $(v_1,...,v_k)\in\sT_mM$ are tangent vectors fixed in a point $m\in U_{\alpha\beta}$, then $\xi_\alpha(v_1,...,v_k)$ and $\xi_\beta(v_1,...,v_k)$ satisfy relation (\ref{diag}). 

Let us assume now that we have a scalar product on $\mg$ which will be denoted by $\langle\cdot,\cdot \rangle_\mg$. In an obvious way it defines a scalar product $K$ on each fiber of the bundle $P\times\mg\to P$ 
$$K:(P\times\mg)\times_P(P\times\mg)\to \mathbb R,\qquad  K(p)(X,Y)=\langle X,Y \rangle_\mg   $$
Let us require now that the form $K$ is $\Ad$-invariant i.e.
$$ K(p)(X,Y)= K(p)(\Ad_gX,\Ad_gY) \qquad g\in G.   $$
In such a case it defines a scalar product $K_A$ in the associated bundle $\ad P\to M$, namely

$$ K_A:\ad P\times_M\ad P\to \mathbb R,\qquad  K_A(m)(X_A(p),Y_A(p))=K(p)(X,Y),\qquad \xi(p)=m.    $$
where $X_A(p)$ and $Y_A(p)$ are equivalence classes of the elements $(p,X)$ and $(p,Y)$ respectively. One can easily see that if $K$ is $\Ad$-invariant the above definition does not depend on the choice of the representative.

Let us assume now that the manifold $M$ is equipped with a metric $g$. It allows us to define a Hodge star operator $\star$. On the other hand the Hodge star allows us to define a scalar product $(\cdot|\cdot)_k$ on a space $\Omega^k(U_\alpha)$
$$(\alpha|\beta)_k=\int\alpha\wedge\star\beta $$

The above scalar product may be extended to a scalar product on a space $\Omega^k(U_\alpha,\mg)$ by 
$$(\cdot|\cdot):\Omega^k(U_\alpha,\mg)\times\Omega^k(U_\alpha,\mg)\to\mathbb R,\quad (\alpha|\beta)=\int K_{ij}\alpha^i\wedge\star\beta^j    $$
where we have used a notation $\alpha=\alpha^i\otimes e_i$, $\beta=\beta^j\otimes e_j$ and $K_{ij}:=K(e_i,e_j)$. In physical literature there is a common notation  $\text{Tr}(\alpha\wedge\star\beta):=K_{ij}\alpha^i\wedge\star\beta^j$. If we assume again that $K$ (we will ommit letter $A$ in $K_A$) is $\ad$-invariant then the above formula defines also a scalar product on a space $\Omega^k(M,\ad P)$.

\subsection{Connection in a principal bundle} 

A connection in a principal bundle $P\to M$ is a $G$-invariant distribution $H$ in $\sT P$ complementary to $\sV P$ i.e.
$$\sT_pP=\mathsf V_pP\oplus H_p,\qquad p\in P$$
and
$$H_pg=H_{pg}  \quad g\in G.$$
The above definition is very elegant and general, however when it comes to applications, it is more convenient to use another definition of connection. We will start with introducing some basic mathematical tools. Let $X$ be an element of $\mg$. The group action of $G$ on $P$ defines a vertical vector field $\sigma_X$ on $P$ associated with the element $X$, namely
$$\sigma_X(p):=\ddt p\exp(tX). $$
The field $\sigma_X$ is called a {\it fundamental vector field} corresponding to an element $X$. The fundamental vector field is equivariant in sense 
$$\sigma_X(pg)=\sigma_{\Ad_{g^{-1}}(X)}(p).$$
The connection in a principal bundle $P$ is a $G$-equivariant, $\mg$-valued one-form $\omega$
$$\omega:\sT P\to\mg, $$
such that $\omega(\sigma_X(p)) = X$ for each $p\in P$ and $X\in\mg$. 
The $G$-equivariance means that 
$$R^*_g\omega(p)=\Ad_{g^{-1}}\circ\omega(p) .$$
Since the connection form is an identity on vertical vectors the difference of two connections is a horizontal form. It follows that the space of connections is an affine subbundle $\mathcal A\subset\sT^*P\otimes\mg$ over $M$ modeled on a vector bundle of $\mg$-valued horizontal forms on $P$. Due to the transformation properties, the bundle of horizontal forms may be identified with the bundle $\sT^*M\otimes\ad P\to M$. We have a diagram
$$\xymatrix{
\mathcal A\subset\sT^*P\otimes\mg  \ar[d]^{pr_P}    \\
 P     \ar@/^1pc/[u]^{\omega} 
} 
$$ 
The connection $\omega$ provides also a decomposition  of the dual bundle $\sT^*P$. The annihilator $(\sV P)^o$ of the bundle $\sV P$ is a vector subbundle in $\sT^*P$ and by definition it is a bundle of  horizontal forms on $P$. The complementary subbundle of $(\sV P)^o$ in $\sT^*P$ is an annihilator of the horizontal distribution $H^o$. Therefore, we have a Whitney sum
$$\sT^*P=(\sV P)^o\oplus_MH^o.$$
We have identifications $(\sV P)^o\simeq P\times\sT^*M$ and $H^o\simeq\sV^*P$. Decomposition of the cotangent bundle allows us to define a horizontal projection of differential forms. If $\beta\in\Lambda^kP$ is a $k$-form on $P$ then its horizontal part is given by a map
$$^h:\Lambda^kP\to\Lambda^k_1P,\quad \beta\longmapsto\beta^h $$ 
$$ \beta^h(v_1,..,v_k)=\beta(v^h_1,..,v^h_k) ,\quad v_1,..,v_k\in\sT P $$
where $v^h_i$ is a horizontal part of the tangent vector $v_i$.

The curvature of the connection is defined as a horizontal part of the differential $\d\omega$. It is a $\mg$ valued two form on $P$ and it reads
$$\Omega=\d\omega+\frac{1}{2}[\omega,\omega],   $$ 
where $[\omega,\omega]$ is a bracket of $\mg$ valued two forms on $P$.

\subsection{Gauge fields}

The notion of the connection may be equally expressed in terms of a family of $\mg$-valued one forms on $M$. This approach is widely used by physicists working in classical field theory. 

Let $s_\alpha:M\supset U_\alpha\to P$ be a local section of the bundle $\pi$. A pull-back of the connection form $\omega$ defines a form $A_\alpha:=s_\alpha^*\omega$
\begin{equation}\label{gaugefield}
A_\alpha(m):\sT_mM\to\mg,\quad v_m\longmapsto\omega_P(s_\alpha(m))(\sT s_\alpha(v_m)). 
\end{equation}
The form $A_\alpha$ is called a {\it gauge field}. In local coordinates it takes a form
$$A_\alpha(m)=A_\alpha^a(m)\otimes e_a=A\ind{^a_j}(m)\mathsf dq^j\otimes e_a  \quad m\in M$$
where we skipped index $\alpha$ to simplify the notation. One can show that gauge fields must satisfy a gluing condition
$$A_\alpha=\Ad_{g_{\alpha\beta}}\circ(A_\beta-g^*_{\alpha\beta}\theta)   $$
or equivalently
$$  A_\alpha=\Ad_{g_{\alpha\beta}}\circ A_\beta+g^*_{\beta\alpha}\theta   $$
on double overlaps $U_{\alpha\beta}$. We used here a Maurer-Cartan form
$$\theta:G\to\sT^*G\otimes\mg,\quad \theta(g)=\sT L_{g^{-1}}.   $$
Each gauge field is a section of the affine subbundle $\mathcal A$ in a tensor bundle $\sT^*M\otimes\mg$
$$\xymatrix{
\mathcal A \subset\sT^*M\otimes\mg  \ar[d]^{pr_M}    \\
 M     \ar@/^1pc/[u]^{A_\alpha} 
}
$$
The bundle of gauge fields $\mathcal A$ is modeled on a vector bundle $\sT^*M\otimes\ad P$.

On the other hand, once we have a family of forms $A_\alpha$ we can restore a connection form $\omega$ on $P$. The cut of $\omega$ to $\pi^{-1}U_\alpha$ is given by
\begin{equation}\label{cut}
\omega_{\alpha}(p)=\Ad_{{g_\alpha(p)}^{-1}}\circ\pi^*A_\alpha(p)+g^*_\alpha\theta(p). 
\end{equation}
The proof of the above statement is rather technical so we will skip it here.

We can also use $s_\alpha$ to pull-back a curvature form $\Omega$ and obtain a $\mg$ valued two-form $F_\alpha=s^*_\alpha\Omega$ on $M$. It is an easy task to check that $F_\alpha$ and $F_\beta$ satisfy a condition (\ref{ad}) on double overlaps. Therefore we claim that the family of forms $\{F_\alpha \}$ define a section 

$$ F:M\to\wedge^2\sT^*M\otimes_M\ad P.$$  
It turns out that the curvature of the connection $\omega$ may be equally represented by a global $\mg$-valued two-form $\Omega$ on $P$ or by a global $\ad P$-valued two-form $F$ on $M$.

\section{Gauge fields in terms of jet spaces}\label{gaugejet}

In gauge field theories the Lagrangian usually depends just on the curvature instead of the full jet of the connection. Therefore it is a natural physical question whether there is a geometrical procedure describing a reduction of the jet of the connection to the curvature. The answer turns out to be "yes" and it is associated with the notion of the so-called {\it canonical principal connection} \cite{Gar1,Gar2,SG}. We will start with a more precise description of the bundle $C\to M$.

\subsection{Jet bundle of a principal bundle}

We will start with a bundle of first jets over a principal bundle. Let us take a closer look on trivialisation of a bundle $\sJ P$. From the previous considerations we see that $\sJ P$ is an affine subbundle in a vector bundle $\sT^*M\otimes_P\sT P$ over $P$. Let $g_\alpha:P\supset \pi^{-1}U_\alpha\to G$ be a local trivialisation of $P$. We can use it to trivialise the tensor bundle $\sT^*M\otimes_P\sT P$, namely

$$\sT^*M\otimes\sT P\to \sT^*M\otimes(\sT M\times G\times\mg)= {\blue(\sT^*M\otimes\sT M)}\times G\times{\red (\sT^*M\otimes\mg)}.$$
The affine subbundle $\sJ P$ in above trivialisation reads that the blue factor is an identity on $\sT M$. Therefore, we can skip it and introduce a local trivialisation
$$\Psi_\alpha:\sJ P\supset\pi_1^{-1}U_\alpha \to  G\times (\sT^*M\otimes\mg),   $$
$$\Psi_\alpha(\sj_m\phi)\longmapsto \Big(g_\alpha(\phi(m)),-\phi^*g_\alpha^*\theta\Big),$$
where $\theta$ is a Maurer-Cartan form on $G$. It can be written as $\theta=\theta^b\otimes e_b$, where $\theta^b$ are one-forms on $\sT^*G$ and $\{e_b\}$ is a basis in $\mg$. Then we obtain a more convenient form of above trivialisation $-\phi^*g_\alpha^*\theta=f^b\otimes e_b$, where $f^b=-\phi^*g_\alpha^*\theta^b$ are one forms on $M$.

Let $g_\alpha:P\supset\pi^{-1}U_\alpha\to G$, $g_\beta:P\supset\pi^{-1}U_\beta\to G$ be two trivialisations. One can show that on double overlaps the transition funtions must obey a gluing condition
$$  -\phi^*g_\beta^*\theta =  -\phi^*g_\alpha^*\theta-\Ad_{g_\alpha\circ\phi(m)^{-1}}\circ g_{\beta\alpha}^*\theta.   $$
The right action $R_g$ of $G$ on $P$ may be prolonged to the action on a first jet bundle $\sJ P$
$$\sJ R_g:\sJ P\to \sJ P: \sj_m\phi\to \sj_m(\phi g),  $$
such that the diagram
$$\xymatrix{
\sJ P   \ar[r]^{\sJ R_g}  \ar[d]_{\pi_{1,0}}  & \sJ P   \ar[d]^{\pi_{1,0}}  \\
P  \ar[r]^{R_g} &  P
} 
$$ 
is commutative. In a local trivialisation above action reads 

$$\sJ R_h: G\times (\sT^*M\otimes\mg) \to  G\times (\sT^*M\otimes\mg)      $$
$$(g, j_\alpha)\longmapsto (gh,\Ad_{h^{-1}}\circ j_\alpha), \qquad j_\alpha\in\sT^*M\otimes\mg  $$

The affine dual of the bundle $\sJ P$ is a bundle $\sJd P$ over $M$. We have a trivialisation
$$\sJd P\to G\times\mathbb R\times\sT M\otimes\mg^*.$$
We introduce coordinates $(g_\alpha(p),r, x^i,\varphi\ind{^a_j})$ in $\sJd P$. If $( e_*^a)$ is a basis in $\mg^*$ then $\varphi\ind{^a_j}$ are coordinates with respect to the basis $\d x^j\otimes e_*^a$. If in a local trivialisation $\sj_m\phi=(g_\alpha\circ\phi(m),x^i, A\ind{^a_j})$ and $T_m=(g_\alpha(p),r, x^i,\varphi\ind{_a^j})$, then the evaluation between $\sJ P$ and $\sJd P$ is given by
$$\langle \sj_m\phi, T_m\rangle=r+\varphi\ind{_a^j}A\ind{^a_j}.  $$
The bundle $\mathsf J^oP$ is locally isomorphic to $G\times\sT M\otimes\mg^*$. It has adapted coordinates $(g_\alpha(p),x^i,\varphi\ind{_a^j})$.

\subsection{Connection in a principal bundle as a section of the jet bundle}

We will present now an equivalent approach to connection in principal bundle. Let $\pi_{1,0}:\sJ P\to P$ be a bundle of first jets over $P$. Let us also consider a section $\sigma:P\to \sJ P$. For each $p\in P$ the value $\sigma(p)$ is a first jet of a certain section of the bundle $\pi$. Let us denote it by $\sigma(p)=\sj_m\phi$ where $\phi(m)=p$. There is a tangent map $\sT\phi:\sT_mM\to\sT_{\phi(m)}P$ associated with $\sj_m\phi$. The image $\sT\phi(\sT_mM)\subset\sT_{\phi(m)}P$ is a horizontal complement to $\sV_mP$ in a tangent space $\sT_{\phi(m)}P$, namely
$$\sT_{\phi(m)}P=\sV_{\phi(m)}P\oplus\sT\phi(\sT_mM).  $$
Therefore, a global section of the bundle $\pi_{1,0}$ provides a horizontal distribution in a bundle $\pi:P\to M$.

Since a principal connection has to be equivariant with respect to the action of $G$ we can divide the bundle $\pi_{1,0}:\sJ P\to P$ by this action and obtain a bundle 
$$\pi_C:C\to M  $$ 
where $C:=\sJ P/G$. Each section of $\pi_C$ provides a principal connection in a bundle $\pi: P\to M$. The bundle $\pi_C$ is an affine bundle over $M$ modeled on a vector bundle $\sT^*M\otimes_P\ad P$. Trivialisation of $C$ comes from the trivialisation of $\sJ P$. We have 
$$C\to\sT^*M\otimes\mg:  [\sj_m\phi]\to (j_\alpha).  $$
The transition functions between different trivialisations read
$$\xymatrix{
       &C    \ar[dr]^{triv_2}   \ar[dl]_{triv_1} &             &        & [\sj_m\phi]      \ar[dr]^{triv_2}   \ar[dl]_{triv_1}   &         \\   
\sT^*M\otimes\mg   \ar[rr] & & \sT^*M\otimes\mg      &    (j_\alpha)   \ar[rr] & & ( \Ad_{g_{\beta\alpha}(m)}\circ j_\alpha+g_{\alpha\beta}^*\theta  )      \\
}
$$
 
There is a one-to-one correspondence between sections of the bundle $\pi_C$ and $G$-invariant sections of the bundle $\pi_{1,0}$ given by the formula
$$ [\sigma]:M\to C, \quad [\sigma](m)=[\sigma(p)], \quad \sigma: P\to\sJ P  $$
where we assume that $\sigma$ is $G$-invariant and $\pi(p)=m$. A glance at the expression shows that above definition does not depend on the choice of the representative.

The $\mg$-valued one forms  $j_\alpha$ may be identified with gauge fields. Therefore, we can introduce coordinates $(x^i,j\ind{^a_k})$ in $C$ such that
$$j_\alpha=j\indices{^a_k}\d x^k\otimes e_a.  $$ 
Since forms $j_\alpha$ may be identified with gauge fields, from now we will rather write $A_\alpha$ and  $A\indices{^a_k}$ instead of $j_\alpha$ and $j\indices{^a_k}$.

The bundle $\sJ C$ may be equipped with local coordinates $(x^i,A\indices{^a_k}, A\indices{^a_{kl}})$. Similary in the dual bundle $\sJd C$ we have coordinates $ (x^i,A\indices{^a_k},r, \varphi\indices{^a_{kl}})$. In the bundle $\sJd C$ we can distinguish a vector subbundle of constant affine maps on each fiber. In coordinates it is given by $(x^i,A\indices{^a_k},r,0)$. Since each such a map depends only on its projection on $C$ it can be associated with a function $f:C\to\mathbb R$. We introduce a bundle

$$\mathsf J^oC:=\sJd C/\{f:C\to\mathbb R\}   $$  
with coordinates $(x^i,A\indices{^a_k},\varphi\indices{^a_k_l})$.

\subsection{Canonical principal connection}

It is well-known that if $P\to M$ is a principal bundle, then there is no canonical choice of the connection in it. However, it turns out that in a case of the bundle $P\times C\to C$ such a canonical choice exists. As it was stated before there is a natural embedding 
$\sJ P\subset\sT^*M\otimes_P\sT P$. What is more, there also exist an embedding of $\sJ P$ in $\sT^*P\otimes_P\sV P$ given by a formula
$$\sJ P\to\sT^*P\otimes\times_P \sV P, \quad \sj_m\phi\longmapsto (id_{\sT P}-\sT(\phi\circ\pi)).   $$
This embedding in an obvious way implies a map
$$ \theta:\sJ P\times_P\sT P\to\sV P, \quad (\sj_m\phi,v)\longmapsto (id_{\sT P}-\sT(\phi\circ\pi))(v),   $$ 
which can be divided by $G$
$$\theta_G:C\times_M\mathsf A\to\ad P,$$
$$(x^i, A\ind{^a_j})\times(x^i, \dot x^k, X^b)\longmapsto (x^i, X^b- A\ind{^a_j}\dot x^j).  $$
The connection in the bundle $P\times C\to C$ is given by a splitting of a short exact sequence
$$0\to\sV(P\times_MC)\to\sT(P\times_MC) \to P\times_M\sT C\to 0  $$
We can rewrite it a bit by using relations 
\begin{eqnarray}
\sV(P\times_MC)&\simeq& C\times\sV P, \\
\sT(P\times_MC)&\simeq& \sT P\times_M\sT C,
\end{eqnarray}
and dividing it by $G$. We obtain then a short exact sequence
$$0\to C\times_M\ad P\to\mathsf A\times_M\sT C \to\sT C\to 0.  $$
One can see that the embedding $\theta_G$ provides a splitting of this sequence. 
$$
\xymatrix{
 0   \ar[r]  & C\times_M\ad P   \ar[r]    &\mathsf A\times_M\sT C \ar[r] \ar[d]^{id\times\tau_{C}} &\sT C \ar[r]  & 0. \\
&   & {\blue   \mathsf A\times_MC    \ar[ul]^{\theta}  }&
}
$$
It follows that if $P\to M$ is a principal bundle then the bundle $P\times_MC\to C$ admits a canonical connection. A form $\theta_G$ defines a horizontal lift
$$Z_C:\sT C\to \sT C\times_M\mathsf A.$$
$$(x^i, A^a_j, \dot x^k, \dot A^b_l)\longmapsto (x^i, A^a_j, \dot x^k, \dot A^b_l)\times (x^i, \dot x^j, A^a_k\dot x^k), $$
which is a section of the tensor bundle 
$$Z_C:C\to\sT^*C\otimes(\sT C\times_M A),  $$
$$Z_C= \d x^i\otimes\partial_{x^i}+\d A^a_j\otimes\partial_{A^a_j}+ A^a_j\d x^j\otimes e_a. $$
We can define now the strength $F_C$ of the canonical connection $Z_C$
$$F_C=\frac{1}{2}[Z_C,Z_C]_{FN}, \quad F_C\in  \wedge^2\sT^*C\otimes_M\ad P  $$
$$F_C= (\d A\ind{^a_j}\wedge\d x^j+\frac{1}{2}c\ind{^a_m_n}\d x^m\wedge\d x^n)\otimes e_a $$
where $[\cdot,\cdot]_{FN}$ is a Frölicher–Nijenhuis bracket. The strength form $F_C$ is called a {\it canonical strength} because if $\omega:M\to C$ is any principal connection then $\omega^*F_C=F_{\omega}$. The form $F_C$ allows us to define an affine surjection
$$\mathcal F:\sJ C\to C\times\wedge^2\sT^*M\otimes_M\ad P, \qquad\mathcal F(\sj\omega)=\omega^*F_C=F_\omega.  $$
There exist a canonical splitting over $C$ \cite{SG}
$$\sJ C=C_+\oplus_CC_-,$$
$$C_+={\mathsf J^2}P/G,    \quad C_-=C\times_M\wedge^2\sT^*M\otimes_M\ad P  $$
and the projections
$$\mathcal F:\sJ C\to C_-,   $$
$$(x^i, A\ind{^a_j},  A\ind{^b_j_k})\longmapsto \Big( x^i, A\ind{^a_j}, \frac{1}{2}(A\ind{^l_j_k}-A\ind{^l_k_j}+c^l_{ab}A\ind{^a_j}A\ind{^b_k}) \Big),  $$
$$\mathcal S:\sJ C\to C_+,  $$
$$(x^i, A\ind{^a_j},  A\ind{^b_j_k})\longmapsto \Big( x^i, A\ind{^a_j}, \frac{1}{2}(A\ind{^l_j_k}+A\ind{^l_k_j}-c^l_{ab}A\ind{^a_j}A\ind{^b_k}) \Big). $$
One can see that the result of the first projection is just a curvature form of the connection
$$F_\omega=\d\omega+\frac{1}{2}[\omega,\omega],$$
$$F_\omega=\frac{1}{2}F\ind{^a_i_j}\d x^i\wedge\d x^j\otimes e_b   $$
where
$$F\ind{^a_i_j}=\partial_jA\ind{^a_i}-\partial_iA\ind{^a_j}+c^a_{mn}A\ind{^m_i}A\ind{^n_j}.$$
Therefore we introduce coordinates 
\begin{eqnarray}
(x^i, A\ind{^a_j}, F\ind{^b_j_k}) &in& C_-,  \\
(x^i, A\ind{^a_j}, S\ind{^b_j_k}) &in& C_+. 
\end{eqnarray}
such that $F\ind{^b_j_k}=-F\ind{^b_k_j}$ and $S\ind{^b_j_k}=S\ind{^b_k_j}$.

\subsection{The phase bundle}

In the previous section we described a procedure of reducing the bundle $\sJ C$ to the bundle $\wedge^2\sT^*M\otimes_M\ad P$. We will show now that $\mathcal F$ defines a similiar reduction on a dual side. We recall that the bundle $\sJd C$ consists of affine maps on $\sJ C$. The bundle $\sJ C$ can be decomposed on subbundles $C_+$ and $C_-$ where the first one is affine and the second one is of vector type. Therefore each affine map on $\sJ C$ may be similiarly decomposed on an affine map on $C_+$ and a linear map on $C_-$. Let us recall that the bundle $\sJd C$ is isomorphic to $\Lambda^{n+1}_2 C$ via the formula
$$\Psi:\Lambda^{n+1}_2 C\to\sJd C  $$
$$\langle \Psi(\alpha),  \sj\omega\rangle\eta=\omega^*\alpha, \qquad  \alpha\in\Lambda^{n+1}_2 C  $$
The pull-back $\omega^*\alpha$ in coordinates reads
$$\alpha(x^i, A\ind{^a_j},p,p\ind{^i^k_a})=p\d^{n+1}x+p\ind{^i^k_a}\d A\ind{^a_k}\wedge\d^{n}x_i,\qquad \omega(x^i)=(x^i,A\ind{^a_k}(x^i))   $$
$$\omega^*\alpha=p\d^{n+1}x+p\ind{^i^k_a}\d A\ind{^a_k}(x^i)\wedge\d^{n}x_i=p\d^{n+1}x+p\ind{^i^k_a}\partial_lA\ind{^a_k}(x^i)\d x^l\wedge\d^{n}x_i $$ 
$$=p\d^{n+1}x+p\ind{^l^k_a}\partial_lA\ind{^a_k}(x^i)\d^{n+1}x= (p+p\ind{^l^k_a}\partial_lA\ind{^a_k}(x^i))\d^{n+1}x.$$
Therefore the evaluation between $\sJ C$ and $\sJd C$ in local coordinates reads
$$\langle \Psi(\alpha),  \sj\omega\rangle=p+p\ind{^l^k_a}A\ind{^a_k_l}. $$
In the same way we can show that the evaluation between $\sJo C\simeq\Lambda^{n+1}_2C/\Lambda^{n+1}_1C$ and $\sJ C$ reads 
$$\langle \Psi(\alpha),  \sj\omega\rangle=p\ind{^l^k_a}A\ind{^a_k_l}. $$
Let us recall that the coordinates $F\ind{^b_j_k}$ and $S\ind{^b_j_k}$ in $C_-$ and $C_+$ are antisymmetric and symmetric in $j,k$ respectively. It means that the dual bundle to $C_-$ in $\sJo C$ is a subbundle $C^*_-\subset\sJo C$ such that each map from $C^*_-$ gives zero on $C_+$. Smilarly, the dual bundle of $C_+$ is a subbundle $C^*_+\subset\sJo C$ such that each map from $C^*_+$ gives zero on $C_-$. We have a decomposition 
$$\sJo C=C_+^*\oplus_CC_-^*  $$
$$ p\ind{^i^k_a}= \frac{1}{2}(p\ind{^i^k_a}+p\ind{^k^i_a})+\frac{1}{2}(p\ind{^i^k_a}-p\ind{^k^i_a})     $$
and coordinates
\begin{eqnarray*}
(x^i, A\ind{^a_j}, \varphi\ind{^i^k_a}) &\text{in}& C^*_-, \qquad \varphi\ind{^i^k_a}=-\varphi\ind{^k^i_a}\\
(x^i, A\ind{^a_j}, \xi\ind{^i^k_a}) &\text{in}&  C^*_+, \qquad \xi\ind{^i^k_a}=\xi\ind{^k^i_a} .
\end{eqnarray*}
One can show that there is an isomorphism 
$$C_-^*=C\times(\wedge^2\TM\otimes_M\ad^*P).  $$
Given a volume form $\eta$ on $M$ we can similarly decompose a bundle   
$$\sJd C=C_+^*\oplus_CC_-^*\oplus\mathbb R.  $$
Since the bundle $\sJd C$ is isomorphic to the bundle $\Lambda\ind{^n_2}(C)$, then each element of $C_-^*\oplus\mathbb R$ may be represented by
$$\alpha(x^i, A\ind{^a_j},p,\varphi\ind{^i^k_a})=p\d^{n+1}x+\varphi\ind{^i^k_a}\d A\ind{^a_k}\wedge\d^{n}x_i$$
Similarly we can identify elements of $C_+^*$ and $C_-^*$ with forms
\begin{eqnarray*}
\Lambda^{n+1}_2C/\Lambda^{n+1}_1C\ni\quad \varphi\ind{^i^k_a}\d A\ind{^a_k}\wedge\d^{n}x_i &\text{in}& C^*_-,\\
\Lambda^{n+1}_2C/\Lambda^{n+1}_1C\ni\quad \xi\ind{^i^k_a}\d A\ind{^a_k}\wedge\d^{n}x_i &\text{in}&  C^*_+.
\end{eqnarray*}
Let us notice in the end that there is a decomposition 
$$\Lambda^{n+1}_2(\sJd C)=\Lambda^{n+1}_2(C^*_+\oplus C^*_-\oplus\mathbb R)=\Lambda^{n+1}_2(C^*_+\oplus\mathbb R)\oplus\Lambda^{n+1}_2(C^*_-\oplus\mathbb R)$$
which means that the canonical multisymplectic form on $\sJd C$ may be written as
$$\Omega=\Omega_++\Omega_-, \qquad \text{where} \quad \Omega_+\in\Lambda^{n+1}_2(C^*_+\oplus\mathbb R), \quad \Omega_-\in\Lambda^{n+1}_2(C^*_-\oplus\mathbb R)   $$ 
and
$$\Omega_-=-\d p\wedge\d^{n+1}x-\d\varphi\ind{^j^k_a}\wedge\d A\ind{^a_j}\wedge\d^{n}x_k.$$
When can easily see that submanifold $(C^*_+\oplus\mathbb R,\Omega_-)$ is still a multisymplectic manifold.
In a case when Hamiltonian section is a section $h:C^*_-\to C^*_-\oplus\mathbb R$ the pull-back of the multisymplectic form reads 
$$h^*\Omega=h^*\Omega_-=\Omega_{-h}\in\Lambda^{n+1}_2(C^*_-).$$

\section{Hamiltonian formalism for gauge theories}\label{Ham}

We will show now Hamiltonian formalism for field theories for which configuration space is $C_-$ instead of full $\sJ C$. For example, this is a case of Yang-Mills theories. Since the bundle of gauge fieds is a bundle $\pi_C:C\to M$, then the role of the configuration space is played by a bundle of first jets $\sJ C$. Therefore, Lagrangian is a map
$${\mathcal L}:\sJ C\to\Lambda^{n+1}M.   $$
Let us assume now that $L$ does not depend on the full jet of connection but only on its projection on $C_-$. Then we have
$$\mathcal L:C\times_M\wedge^2\sT^*M\otimes_M\ad P\to\Lambda^{n+1}M.   $$
In a presence of a volume form the Lagrangian density can be written as $\mathcal L=L\eta$, where $L$ is called {\it Lagrangian function}. In coordinates we have
$$ \mathcal L(x^i, A\ind{^a_j}, F\ind{^b_j_k})=L(x^i, A\ind{^a_j}, F\ind{^b_j_k})\eta.   $$
The Hamiltonian side of the theory is usually obtained from the Lagrangian density $\mathcal L$ via Legendre transform. Since the Lagranian depends only on $C_-$ and 
$$A\ind{^b_j_k}= F\ind{^b_j_k}+ S\ind{^b_j_k}   $$
$$\frac{\partial}{\partial A\ind{^b_j_k}}=\frac{\partial}{\partial F\ind{^b_j_k}}+\frac{\partial}{\partial S\ind{^b_j_k}}   $$
the Legendre map $\sJ C\to\sJd C$ reduces to
\begin{equation}\label{Leg1}
{\it Leg}_{\mathcal L}:C_-\to C^*_-\oplus\mathbb R, \\
\end{equation}

\begin{equation}\label{Leg2}
{\it Leg}_{\mathcal L}(x^i, A\ind{^a_j}, F\ind{^b_j_k})=\left( x^i, A\ind{^a_j}, L- \frac{\partial L}{\partial F\ind{^b_j_k}}\\ F\ind{^b_j_k}, \frac{\partial L}{\partial F\ind{^b_j_k}}  \right)
\end{equation}
Using the projection $\mu:\sJd C\to \sJo C$ we can now introduce a map 
\begin{equation}\label{leg1}
{\it leg}_{\mathcal L}=\mu\circ{\it Leg}_{\mathcal L},\quad {\it leg}_{\mathcal L}:C_-\to C^*_-,   
\end{equation}

\begin{equation}\label{leg2}
{\it leg}_{\mathcal L}(x^i, A\ind{^a_j}, F\ind{^b_j_k})=\left( x^i, A\ind{^a_j}, \frac{\partial L}{\partial F\ind{^b_j_k}} \right).
\end{equation}
Hamiltonian section is then defined as
$$h:C^*_-\to C^*_-\oplus\mathbb R,\quad   h={\it Leg}_{\mathcal L}\circ {\it leg}_{\mathcal L}^{-1},$$
$$h(x^i, A\ind{^a_j}, \varphi\ind{_b^j^k})=(x^i, A\ind{^a_j}, L(x^i, A\ind{^a_j}, F\ind{^b_j_k})-\varphi\ind{_b^j^k}F\ind{^b_j_k} , \varphi\ind{_b^j^k})   $$
with the associated Hamiltonian density
$$\mathcal H:C^*_-\oplus\mathbb R\to\Omega^{n+1}(M),  $$
$$\mathcal H(x^i, A\ind{^a_j}, p, \varphi\ind{_b^j^k})=(p+ \varphi\ind{_b^j^k}F\ind{^b_j_k}-L)\d^nx $$ 
and the Hamiltonian function
\begin{equation}\label{Ham}
H(x^i, A\ind{^a_j}, \varphi\ind{_b^j^k})= \varphi\ind{_b^j^k}F\ind{^b_j_k}-L.
\end{equation}
Let us consider now a section 
$$\tau:M\to C^*_-, \quad \tau(x^i)=\Big(x^i,\tau\ind{^a_j}(x^i),\tau\ind{_a^j^k}(x^i)\Big).  $$
The Hamilton-De Donder-Weyl equations on $C^*_-$ read
$$\frac{\partial\tau\ind{^a_j}}{\partial x^i}=\frac{\partial H}{\partial \varphi\ind{_a^i^j}}\circ\tau, \qquad  
\frac{\partial\tau\ind{_a^i^j}}{\partial x^i}=-\frac{\partial H}{\partial A\ind{^a_j}}\circ\tau$$

\section{Hamilton-Jacobi theory for gauge theories}

In this section we will apply a formalism derived in \cite{HJ1} to the case of gauge fields. Let us consider a diagram
$$\xymatrix{
\mathsf J^oC      \ar[dr]^{\pi^o_1}    \\
 \sJd C  \ar[d]   \ar[u]^{\mu} & M  \\
C   \ar@/^/[u]^{\gamma} \ar[ur]_{\pi} 
}
$$
where $\gamma$ is a section. We will assume now that we have a connection in a bundle $\sJo C\to M$ and ${\bf h}$ is its horizontal projector. The connection in $\pi_{1}^o$ may be projected to a connection in a bundle $C\to M$ in such a way that its horizontal projector reads 
$${\bf h}^\gamma(\omega):\sT_\omega C\to\sT_\omega C,$$
$$ X\longmapsto {\bf h}^\gamma(\omega)(X)=\sT_f\pi^\circ_{1,0}({\bf h}(f)(Y)),  $$
where $f=(\mu\circ\gamma)(\omega)$ and $Y$ is any vector field of $\sT_f\sJo C$ which projects onto $X$ by $\sT\pi^\circ_{1,0}$. The idea of the Hamilton-Jacobi equation can be expressed as follows.

\begin{theorem}
Assume that $\gamma$ is closed and that the induced connection $\bf{h}^\gamma$ on $C\to M$ is flat. Then the following conditions are equivalent: \\
i) If $\sigma$ is an integral section of $\bf h$ then $\mu\circ\gamma\circ\sigma$ is a solution of the Hamilton's equations   \\
ii) The $(n+1)$-form $ h\circ\mu\circ\gamma$ is closed 
\end{theorem}
In our case the above diagram may be reduced to the diagram
$$\xymatrix{
  C^*_-   \ar[dr]^{\pi^o_1}  \ar@/_/[d]_{h}   \\
 C^*_-\oplus\mathbb R  \ar[d]   \ar[u]_{\mu} & M  \\
C   \ar@/^/[u]^{\gamma_-} \ar[ur]_{\pi}   \ar@/^3pc/[uu]^{\lambda} 
}
$$
Let us denote by $\lambda:=\mu\circ\gamma_-$ a section of the bundle $C^*_-\to C$. We have the following theorem.
\begin{theorem}
Assume that $\gamma_-$ is closed and that the induced connection $\bf{h}^\gamma$ on $C\to M$ is flat. Then the following conditions are equivalent: \\
i) If $\sigma$ is an integral section of $\bf h$ then $\lambda\circ\sigma$ is a solution of the Hamilton's equations   \\
ii) The $(n+1)$-form $ h\circ\lambda$ is closed 
\end{theorem}
Let us assume now that $\lambda=\d S$ where
$$S=S^i(x^i,A\ind{^a_j})\d^nx_i$$ 
is a $1$-semibasic form on $C$ such that $\frac{\partial S_i}{\partial A\ind{^a_j}}=-\frac{\partial S_j}{\partial A\ind{^a_i}}$. Then in local coordinates the equation $\d(h\circ\lambda)=0$ is equivalent to
\begin{equation}\label{redHJ}
\frac{\partial S^i}{\partial x^i }+H\left(x^i,A\ind{^a_j},\frac{\partial S_i}{\partial A\ind{^a_j}}\right)=f(x^i).
\end{equation}

\section{Example: Yang-Mills theories}

Let us consider now a Hamilton-Jacobi theory for Yang-Mills theory. A free Yang-Mills theory is described by a Lagrangian 
$${\mathcal L}:\sJ C\supset C\times_M\wedge^2\sT^*M\otimes_M\ad P\to\Omega^{n+1},\quad {\mathcal L}(\omega,F)=\frac{1}{4}K_{ab}F^a\wedge\star F^b. $$
where $F=F^a\otimes e_a=F\ind{^a_i_j}\d x^i\wedge\d x^j\otimes e_a$ . One can easily show that the associated Lagrangian function is
$${ L}(q^i, A\ind{^a_j}, F\ind{^b_j_k})=\frac{1}{4}K_{ab}g^{im}g^{jn}F\ind{^a_i_j}F\ind{^b_m_n}\sqrt{|g|}=\frac{1}{4}F\ind{^a_i_j}F\ind{_a^i^j}\sqrt{|g|} $$
where $g$ is a metric on $M$ and $K$ is an $\Ad$-invariant scalar product on $\mg$. To write the Hamilton-Jacobi equation we need a Hamiltonian side of the theory. Using (\ref{Leg2}) and (\ref{leg2}) we have
$${\it leg}_{\mathcal L}(x^i, A\ind{^a_j}, F\ind{^a_i_j})=\left( x^i, A\ind{^a_j}, \frac{\partial L}{\partial F\ind{^a_i_j}} \right)=
\left( x^i, A\ind{^a_j}, \frac{1}{2}F\ind{_a^i^j}\sqrt{|g|} \right) $$

$${\it Leg}_{\mathcal L}(x^i, A\ind{^a_j}, F\ind{^a_i_j})=\left( x^i, A\ind{^a_j}, -\frac{1}{4}F\ind{^a_i_j}F\ind{_a^i^j}\sqrt{|g|}, \frac{1}{2}F\ind{_a^i^j}\sqrt{|g|} \right)$$
The inverse map of ${\it leg_{\mathcal L}}^{-1}$ reads 
$$ {\it leg}^{-1}_{\mathcal L}(x^i, A\ind{^a_j}, \varphi\ind{_a^i^j})=\left( x^i, A\ind{^a_j}, \frac{2}{\sqrt{|g|}}\varphi\ind{^a_i_j} \right). $$
from which we have that
$$F\ind{^a_i_j}=\frac{2}{\sqrt{|g|}}\varphi\ind{^a_i_j}$$
Therefore from (\ref{Ham}) we obtain that the Yang-Mills Hamiltonian reads
$$H(x^i, A\ind{^a_j}, \varphi\ind{_a^i^j})= \frac{2}{\sqrt{|g|}}\varphi\ind{_a^i^j}\varphi\ind{^a_i_j}-L=
\frac{2}{\sqrt{|g|}}\varphi\ind{_a^i^j}\varphi\ind{^a_i_j}-\frac{1}{4}\cdot\frac{4}{|g|}\varphi\ind{_a^i^j}\varphi\ind{^a_i_j}{\sqrt{|g|}}=
\frac{1}{\sqrt{|g|}}\varphi\ind{_a^i^j}\varphi\ind{^a_i_j}$$
We can find now Hamilton-De Donder-Weyl equations
$$\frac{\partial\tau\ind{^a_j}}{\partial x^i}=\frac{\partial H}{\partial \varphi\ind{_a^i^j}}\circ\tau, \qquad  
\frac{\partial\tau\ind{_a^i^j}}{\partial x^i}=-\frac{\partial H}{\partial A\ind{^a_j}}\circ\tau$$ 
of above Hamiltonian. Let us consider a section
$$\tau(x^i)=\Big(x^i,\tau\ind{^a_j}(x^i),\tau\ind{_a^j^k}(x^i)\Big)  $$
The derivatives of the Hamiltonian read
$$\frac{\partial H}{\partial \varphi\ind{_a^i^j}}=\frac{2}{\sqrt{|g|}}\varphi\ind{^a_i_j}    $$
$$\frac{\partial H}{\partial A\ind{^a_j}}=\frac{\partial H}{\partial\varphi\ind{_b^k^l}}\cdot\frac{\partial\varphi\ind{_b^k^l}}{\partial A\ind{^a_j}}=\frac{2}{\sqrt{|g|}}\cdot\varphi\ind{^b_k_l}\cdot\frac{\sqrt{|g|}}{2}\cdot\frac{\partial F\ind{_b^k^l}}{\partial A\ind{^a_j}}=
\varphi\ind{^b_k_l}\cdot\frac{\partial F\ind{_b^k^l}}{\partial A\ind{^a_j}}   $$
where
$$\frac{\partial F\ind{_b^k^l}}{\partial A\ind{^a_j}}=c\ind{_b_m_a}(A\ind{^m^k}\delta\ind{^l_j}-A\ind{^m^l}\delta\ind{^k_j})$$
so that
$$\frac{\partial H}{\partial A\ind{^a_j}}=\varphi\ind{^b_k_l}c\ind{_b_m_a}(A\ind{^m^k}\delta\ind{^l_j}-A\ind{^m^l}\delta\ind{^k_j})= 2\varphi\ind{^b_k_l}c\ind{_b_m_a}A\ind{^m^k}\delta\ind{^l_j}=2\varphi\ind{_b^k^j}c\ind{^b_m_a}A\ind{^m_k}.$$
In above calculations $c\ind{^k_a_b}$ are constant structures of a Lie algebra $\mg$. Therefore, Hamilton-De Donder-Weyl equations for Yang-Mills theory read
$$\frac{\partial\tau\ind{^a_j}}{\partial q^i}=\frac{2}{\sqrt{|g|}}\varphi\ind{^a_i_j}  , $$
$$\frac{\partial\tau\ind{_a^i^j}}{\partial q^i}=-2\varphi\ind{_b^k^j}c\ind{^b_m_a}A\ind{^m_k}.$$ 
We can also write the Hamilton-Jacobi equation 
$$\frac{\partial S_i}{\partial x^i }+\frac{1}{\sqrt{|g|}}K^{ab}g_{ki}g_{lj}\frac{\partial S_i}{\partial A\ind{^a_j}}\frac{\partial S_k}{\partial A\ind{^b_l}}=f(x^i).  $$

\section{Hamilton-Jacobi theory for gauge theories in a Cauchy data space}

The Cauchy data space itself allows to relate a finite-dimensional and an infinite-dimensional formulation of field theory. It was shown in \cite{HJ1} that the Hamilton-Jacobi theory for classical fields may be formulated in a setting of the Cauchy space. We will show here a similar construction of the Hamilton-Jacobi theory for gauge field theories. We will base on the notation and constructions from \cite{HJ1}.

\subsection{A space of Cauchy data}

We will introuce now a notion of a Cauchy data space which is fundamental tool to consider infinite-dimensional version of field theory. Let $M$ be an $n+1$-dimensional manifold and $\Sigma$ an $n$-dimensional, compact, oriented and embeded submanifold of $M$. We say that $\Sigma$ is a {\it Cauchy surface}. Let us assume that $\Sigma$ has a volume form $\eta_\Sigma$ such that
$$\int_\Sigma\eta_\Sigma=1.   $$   
We define a slicing of $M$ which is a diffeomorphism
$$\chi_M:\mathbb R\times \Sigma\to M, \qquad (t,p)\longmapsto\chi_M(t,p).  $$
For each fixed $t\in\mathbb R$ the slicing defines an embedding of $\Sigma$ in $M$
$$(\chi_M)_t:\Sigma\to M, \quad p\longmapsto\chi_M(t,p)   $$ 
We will also assume that there exist such $t_0\in\mathbb R$ that $\Sigma=(\chi_M)_{t_0}(\Sigma)$. The infinitesimal generator of $\chi_M$ will be denoted by $\xi_M$
$$\xi_M:M\to\mathfrak X(M),\quad \xi_M:=(\chi_M)_*\Big(\frac{\partial}{\partial t}\Big)  $$
where $\frac{\partial}{\partial t}$ is a vector field tangent to the curve $(t,x)\to(t+s,x)$. 

Let $E\to M$ be a fibre bundle and let us define a set 
$$\wdt E:=\{\sigma:\Sigma\to E, \quad  \text{such that}\quad  \pi\circ\sigma=(\chi_M)_t  \}.  $$
It is a line bundle over $\mathbb R$ and there is a one-to-one correspondence between sections of $E\to M$ and sections of $\wdt E\to\mathbb R$. If $\phi:M\to E$ is a section than the section of $\wdt E\to\mathbb R$ is given by
$$t\longmapsto \sigma_t=\phi\circ(\chi_M)_t.  $$

Furthermore, the standard bundle structures we know define by composition the bundle structures on the space of $\chi$-sections. For instance one can construct a bundle
$$\wdt{\pi}^o_{1,0}:\wdt{\sJo E}\to\wdt E, \quad \sigma_{\sJo E}\longmapsto\pi^o_{1,0}\circ\sigma_{\sJo E}.   $$ 
One can also define tangent vectors and forms on $\wdt E$. A vector tangent to the curve $t\to\sigma_t$ is given by map a $\wdt X:\Sigma\to\sT E$ such that the diagram
 
$$\xymatrix{
\Sigma  \ar[r]^{\wdt X}  \ar[dr]_{\sigma_0}& \sT E \ar[d]^{\tau_E}   \\
  & E  
}
$$
is commutative and such that there exist such $k\in\mathbb R$ that 
$$\sT\pi(\wdt X(p))=k\xi_M\Big(\pi(\sigma_E(p)) \Big),\quad \forall p\in N   $$
Forms on $\wdt E$ can be constructed from forms on $E$. Let $\alpha$ be a $k$-form on $E$. The form $\wdt\alpha$ on $\wdt E$ is given by 
$$\wdt\alpha(\sigma_E)(\wdt X_1,...,\wdt X_k)=\int_\Sigma\sigma_E^*(i_{\wdt X_1,...,\wdt X_k}\alpha)  $$
In particular we will be interested in a form $\wdt\Omega_h$ obtained from a form $\Omega_h$.

\subsection{The Hamiltonian-Jacobi theory on a bundle $\wdt C$}

Let us consider now a case of $E=C$. The bundle  $\wdt C$ inherits an affine structure from $C\to M$ where the model bundle is a vector bundle $\wdt{\sT^*M\otimes\ad P}\to\mathbb R$. The correspondence between $C\to M$ and $\wdt C\to\mathbb R$ constitutes the relation between the infinite-dimensional and finite-dimensional picture of gauge field theory.

For a section $\tau:M\to\sJo C$ there is an associated section 
$$c:\mathbb R\to\wdt{\sJo C},\qquad t\longmapsto \tau_{|N_t}\circ(\chi_M)_t.  $$
One can show that $\tau$ satisfies Hamiltons equations if and only if 
$$i_{\dot c(t)}\wdt\Omega_h=0.   $$
The above formula is an infinite-dimensional version of the Hamilton's equations on a bundle $C$.

Let us move now to the Hamilton-Jacobi equation. Let us assume now that $\gamma$ is a solution of the Hamilton-Jacobi equation and that we have a connection in a bundle $\sJd C\to M$ which satisfies Hamilton's field equations.

The section $\gamma$ defines a section 
$$\wdt\gamma:\wdt C\to \wdt{\sJo C},  $$
$$\sigma_C\longmapsto\mu\circ\gamma\circ\sigma_C.  $$

On the other hand we can construct vector fields $\widetilde X^\bh$ and $\widetilde X^{\bh^\gamma}$ from the connections $\bh$ and $\bh^\gamma$ by

$$\widetilde X^\bh:\widetilde{\sJo C}\to\sT\widetilde{\sJo C},   $$
$$\sigma_{\sJo}\to\widetilde X^\bh(\sigma)\quad :\Sigma\to\sT\sJo C    $$
$$p\longmapsto \widetilde X^\bh(\sigma)(p)=\text{Hor}\Big(\xi((\chi_M)_t(p))  \Big)  $$
where $\text{Hor}(X)(y)$ represents the horizontal lift of the tangent vector $X$ to the point $y$.

In the same way we can construct the vector field $\widetilde X^{\bh^\gamma}$ on $\widetilde C$ using the horizontal lift of the connection $\bh^\gamma$. Now we can introduce an infinite-dimensional version of the Hamilton-Jacobi theorem (\cite{HJ1}):
\begin{theorem} The section $\wdt\gamma$ satisfies:\\
i) $\wdt{\gamma}^*\wdt{\Omega}_{h}=0$ \\
ii) $i_{\sT\wdt\gamma(\sigma_C)(\wdt X^{h^\gamma})}\wdt\Omega_h=0 \text{ for all } \sigma_C\in\wdt C \text{ which is an integral submanifold of the connection } \bf h^\gamma_{\widetilde \pi(\sigma_C)}. $
\end{theorem}

\subsection{Reduced Hamilton-Jacobi in a space of Cauchy data}

In the previous subsection we presented the infinite-dimensional formulation of Hamilton-Jacobi equation for gauge theories. Now we will show how to reduce it when Hamiltonian depends only on the part $C^*_-$. 

At the beginning let us notice that many of our constructions from section 4 have their reflections in the bundles of $\chi$-sections. For example, the decomposition $\sJ C=C_-\oplus_CC_+$ implies a decomposition $\wdt{\sJ C}=\wdt{C_+}\oplus_{\wdt C}\wdt{C_-}$. The same happens with the other bundles namely $\wdt{\sJo C}=\wdt{C^*_+}\oplus_{\wdt C}\wdt{C^*_-}$ and $\wdt{\sJd C}=\wdt{C^*_+}\oplus_{\wdt C}\wdt{C^*_-}\oplus\mathbb R$.
Let us consider now a cut of $\Omega$ to $C^*_-$. An infinite-dimensional counterpart of $(C^*_-,\Omega_{-h})$ is a pair $(\wdt C^*_-,\wdt\Omega_{-h})$. Let us remind that $(C^*_-,\Omega_{-h})$ is a multisymplectic manifold while $(\wdt C^*_-,\wdt\Omega_{-h})$ is presymplectic.

Let us assume now that $\gamma$ is a solution of the reduced Hamilton-Jacobi equation (\ref{redHJ}). We define a section
$$\wdt\gamma_-:\wdt C\to \wdt{C^*_-}\oplus\mathbb R,  $$
$$\sigma_C\longmapsto\mu\circ\gamma_-\circ\sigma_C.  $$
One can easily see that
$$\wdt{\gamma_-}^*\wdt{\Omega}_{h}=\wdt{\gamma_-}^*(\wdt{\Omega}_{+h}+\wdt{\Omega}_{-h})=\wdt{\gamma_-}^*\wdt{\Omega}_{-h}.  $$
On the other hand ${\sT\wdt\gamma_-(\sigma_C)(\wdt X^{h^\gamma_-})}\in\wdt C^*_-$ so that $i_{\sT\wdt\gamma_-(\sigma_C)(\wdt X^{h^\gamma_-})}\wdt\Omega_{h}=i_{\sT\wdt\gamma_-(\sigma_C)(\wdt X^{h^\gamma_-})}\wdt\Omega_{-h}.$

Hence, the reduced infinite-dimensional Hamilton-Jacobi theory says that the two following statements are equivalent:

\begin{theorem} The section $\wdt\gamma:\wdt C\to \wdt{C^*_-}\oplus\mathbb R$ satisfies:\\
i) $\wdt{\gamma}^*\wdt{\Omega}_{-h}=0$ \\
ii) $i_{\sT\wdt\gamma(\sigma_E)(\wdt X^{h^\gamma})}\wdt\Omega_{-h}=0 \text{ for all } \sigma_E\in\wdt E \text{ which is an integral submanifold of the connection } \bf h^\gamma_{\widetilde \pi(\sigma_C)}. $
\end{theorem}

\section*{Acknowledgements}

This work has been partially supported by MINECO Grants MTM2016-76-072-P and the ICMAT Severo Ochoa projects SEV-2011-0087 and SEV-2015-0554.

\end{document}